\documentclass[lettersize,journal]{IEEEtran}
\usepackage{amsmath,amsfonts}
\usepackage{algorithm}
\usepackage{algpseudocode}
\usepackage{graphicx}
\usepackage{array}
\usepackage[caption=false,font=normalsize,labelfont=sf,textfont=sf]{subfig}
\usepackage{textcomp}
\usepackage{url}
\usepackage{array}
\newcolumntype{C}[1]{>{\centering\arraybackslash}p{#1}}
\usepackage{verbatim}
\usepackage{graphicx}
\usepackage{cite}
\usepackage{amssymb}
\usepackage{amsmath}
\allowdisplaybreaks
\usepackage{mathrsfs}
\usepackage{lipsum} 
\usepackage{booktabs}
\usepackage{afterpage}
\usepackage{enumerate}
\usepackage{cleveref}
\usepackage{stfloats}
\usepackage{graphicx}
\usepackage{subfig}
\usepackage{graphicx}
\usepackage{titlesec}
\usepackage{indentfirst}
\titlespacing*{\section}{3pt}{3pt}{3pt}
\titlespacing*{\subsection}{3pt}{3pt}{3pt}

\renewcommand{\thetable}{\arabic{table}}  
\begin{document}
\renewcommand{\bottomfraction}{1.0} 
\setcounter{bottomnumber}{5}       
\hyphenation{op-tical net-works semi-conduc-tor IEEE-Xplore}
\renewcommand{\topfraction}{1.0} 
\renewcommand{\bottomfraction}{1.0} 
\renewcommand{\textfraction}{0.0} 
\renewcommand{\thefigure}{\arabic{figure}}
\crefformat{figure}{Fig.~#2(#1)#3}
\setcounter{topnumber}{5} 
\setcounter{bottomnumber}{5} 
\setcounter{totalnumber}{10} 


\title{Low-Complexity Channel Estimation in OTFS Systems with Fractional Effects}

\author{Guangyu Lei, Yanduo Qiao, Tianhao Liang,~\IEEEmembership{Member,~IEEE}, Weijie Yuan, ~\IEEEmembership{Senior Member,~IEEE}, \\and Tingting Zhang,~\IEEEmembership{Member,~IEEE}
\thanks{Guangyu Lei, Yanduo Qiao, Tingting Zhang, and Tianhao Liang are with the School of Information Science and Technology, Harbin Institute of Technology (Shenzhen), Shenzhen 518055, China, and also with the Guangdong Provincial Key Laboratory of Space-Aerial Networking and Intelligent Sensing, Shenzhen, China (e-mail: lgy770453058@163.com, 22s152070@stu.hit.edu.cn, liangth@hit.edu.cn, zhangtt@hit.edu.cn). Tingting Zhang is also with Pengcheng Laboratory, Shenzhen 518055, China. Weijie Yuan is with the School of System Design and Intelligent Manufacturing, Southern University of Science and Technology, Shenzhen, China (e-mail: yuanwj@sustech.edu.cn).}}

\markboth{Journal of \LaTeX\ Class Files,~Vol.~14, No.~8, August~2021}%
{Shell \MakeLowercase{\textit{et al.}}: A Sample Article Using IEEEtran.cls for IEEE Journals}


\maketitle

\begin{abstract}
Orthogonal Time Frequency Space (OTFS) modulation exploits the sparsity of Delay-Doppler domain channels, making it highly effective in high-mobility scenarios. Its accurate channel estimation supports integrated sensing and communication (ISAC) systems. The letter introduces a low-complexity technique for estimating delay and Doppler shifts under fractional effects, while addressing inter-path interference. The method employs a sequential estimation process combined with interference elimination based on energy leakage, ensuring accurate channel estimation. Furthermore, the estimated channel parameters can significantly improve ISAC system performance by enhancing sensing capabilities. Experimental results validate the effectiveness of this approach in achieving accurate channel estimation and facilitating sensing tasks for ISAC systems.
\end{abstract}

\begin{IEEEkeywords}
OTFS, channel estimation, low-complexity, inter-path interference, integrated sensing and communication.
\end{IEEEkeywords}

\section{Introduction}

\IEEEPARstart{O}{rthogonal} Time Frequency Space (OTFS) modulation, operating within the Delay-Doppler (DD) domain, exploits the inherent sparsity of DD domain channels to deliver outstanding performance in high-mobility environments. OTFS focuses on channel representation in the DD domain, where its estimation process effectively captures parameters like the range and velocity of reflectors, making it a promising candidate for Integrated Sensing and Communication (ISAC) applications. Recent research has increasingly explored OTFS's potential for ISAC \cite{ref1,ref2,ref3,ref4}. For example, \cite{ref5} demonstrated mutual positioning in UAV formations utilizing OTFS pilots and DD domain information, while \cite{ref6} leveraged OTFS's robustness in high-mobility contexts to ensure reliable communication coverage for high-speed trains through UAV-mounted OTFS devices.

Accurate estimation of path's delay, Doppler shift, and channel gain enables precise reconstruction of DD domain channels, thereby enhancing communication performance. Concurrently, precise parameter estimation of scene targets facilitates augmented system sensing capabilities. However, fractional delay and Doppler effects pose significant challenges to accurate channel estimation and radar parameter sensing. Inter-path interference (IPI), exacerbated by fractional delays and Doppler shifts, further complicates these scenarios by substantially degrading channel estimation and sensing performance. Additionally, a low-complexity approach is essential for real-time channel estimation in dynamic scenarios such as UAV formation communications \cite{ref5}, where channel parameters fluctuate rapidly. Therefore, developing low-complexity and robust IPI elimination strategies is crucial for significantly enhancing channel estimation accuracy and sensing performance.

Previous research has made significant progress in addressing these challenges. For example, \cite{ref7} formulated the fractional OTFS channel estimation problem as a sparse recovery task, successfully obtaining precise estimates by exploiting DD domain sparsity. However, the approach causes high computational complexity, limiting its applicability in dynamic environments. Similarly, a low-complexity maximum likelihood estimation (MLE) method is proposed in \cite{ref8} that avoids matrix inversion but fails to eliminate IPI adequately. Additionally, \cite{ref9} extended the MLE method in \cite{ref8} and developed an iterative IPI elimination scheme; nevertheless, the scheme did not explore the sequential ordering of path estimation and interference elimination, and the iterative process also led to increased algorithmic complexity.

To reduce the complexity, the estimation could be implemented directly on the effective channel. \cite{ref10} provided a threshold-based channel estimation method on the DD domain effective channel, yet it failed to estimate paths' delays accurately, Doppler shifts, and channel gains. Building upon this, \cite{ref11} extended the approach by reconstructing the Doppler response to achieve channel estimation under a fractional Doppler framework through linear correlation, thereby precisely estimating paths' Doppler shifts. However, the method does not adequately address path identification or effectively eliminate IPI.

To address these challenges, the letter proposes a low-complexity path estimation algorithm. 
\begin{enumerate}
\item The path extraction method in \cite{ref10} is improved to identify all potential paths by analyzing the DD domain characteristics of fractional effects.
\item Accurate estimation of delay and Doppler shift is achieved by modeling the delay and Doppler response under the fractional effects.
\item “Energy leakage” is introduced to quantify fractional effects and guide IPI elimination priorities.
\end{enumerate}
The experimental results confirm the algorithm’s effectiveness and accuracy in channel estimation and assisting the sensing performance.

\begin{figure}[H]
\centering
\subfloat{\includegraphics[width=3.5in]{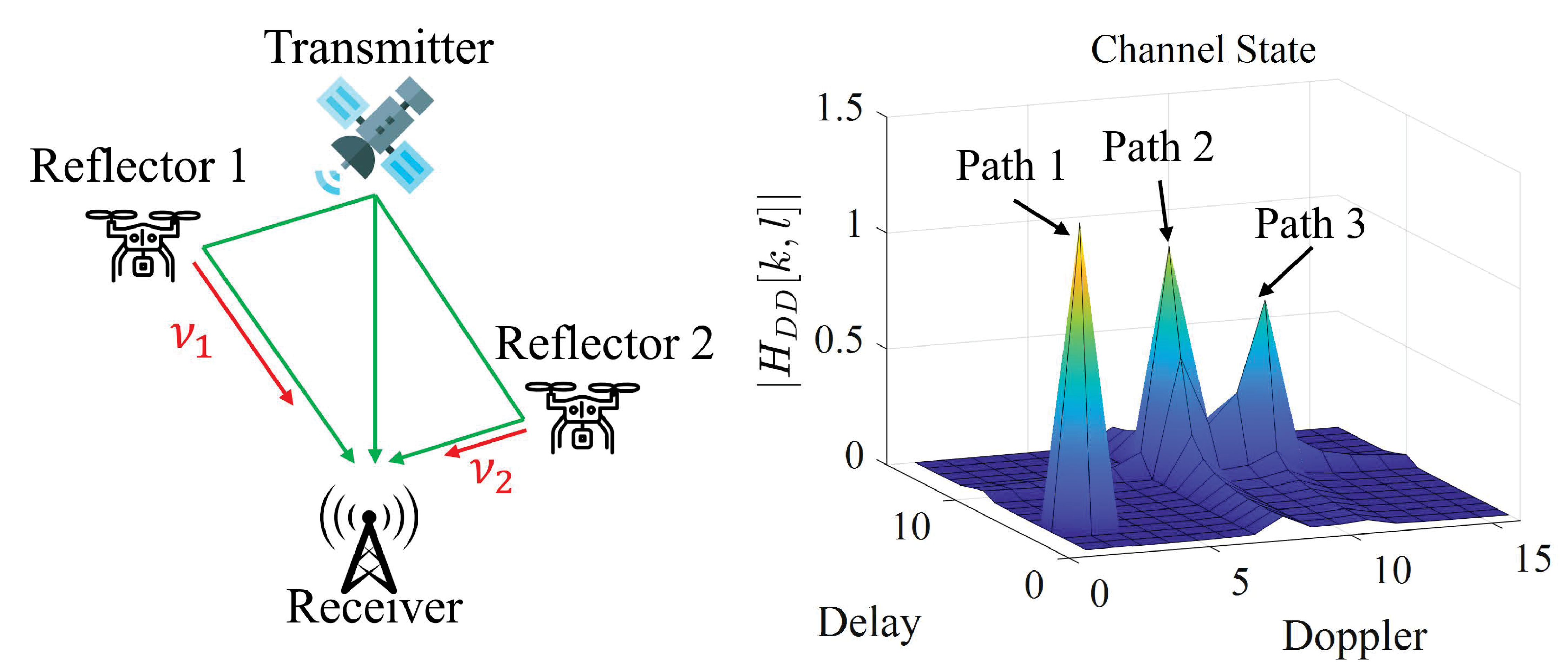}%
\label{fig_first_case}}
\hfil
\subfloat[]{\includegraphics[width=3.5in]{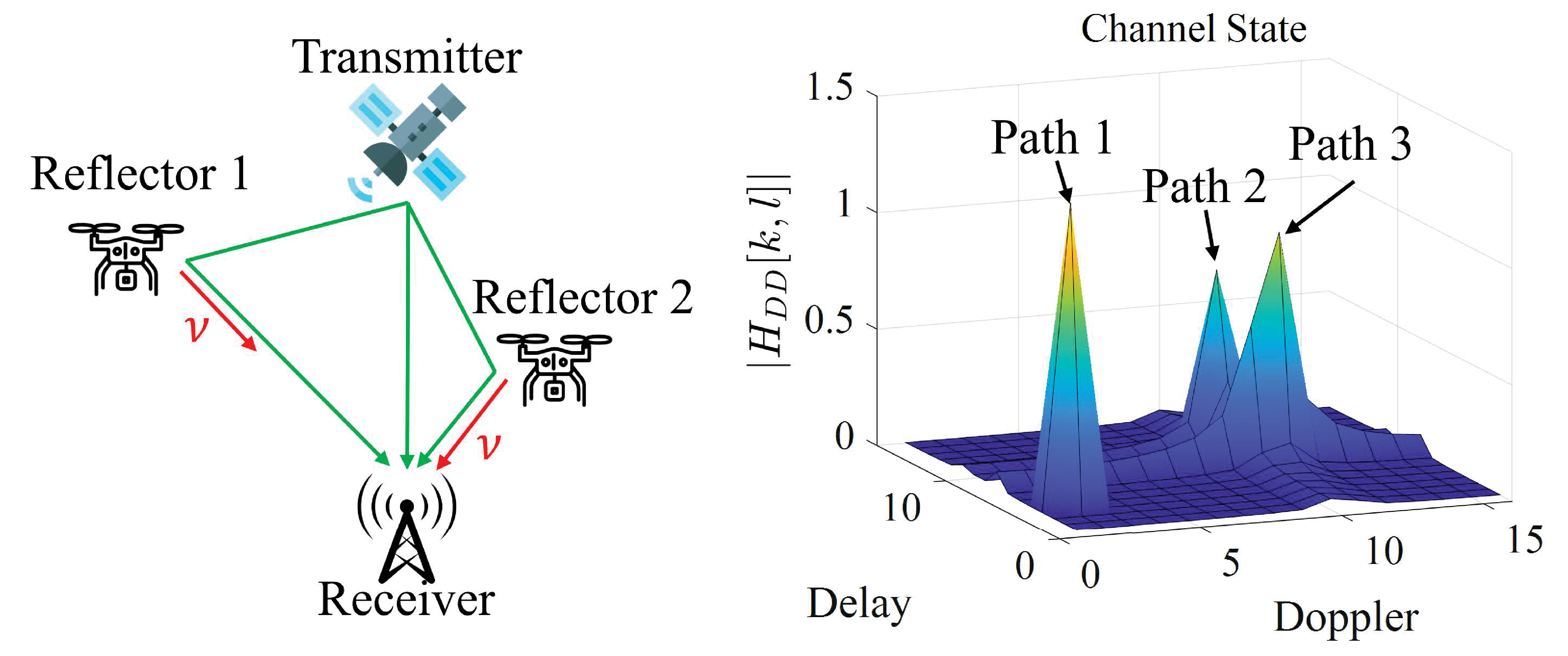}%
\label{fig_second_case}}
\caption{Schematic illustration of OTFS communication assists sensing, where different reflectors' states lead to different channel situations. (a) The channel state when there are two paths with the same delay. (b) The channel state when there are two paths with the same Doppler.}
\label{fig_1}
\end{figure}

\section{System Model}

OTFS signals are modulated in the DD domain, with data symbols mapped onto the delay-Doppler resource element (DDRE) of $ M \times N $. The transmitted signal at each grid point is represented as $ \mathbf{X}[k,l] $, where $ k $ and $ l $ are the Doppler and delay indices ($ k = 0, 1, \dots, N-1 $, $ l = 0, 1, \dots, M-1 $). The grid resolutions in delay and Doppler are $ \frac{1}{M \Delta f} $ and $ \frac{1}{NT} $, corresponding to ranges of $ T $ and $ -\Delta f / 2 $ to $ \Delta f / 2 $, respectively.
 
In practical systems, the delay and Doppler shifts of paths are often not integer multiples of the DDRE grid step sizes but instead include fractional components, referred to as fractional effects. The letter defines the delay and Doppler values of the $ p $-th path as $ \tau_p $ and $ \nu_p $, where $  \nu_p =\frac{k_{\nu_{p}}\Delta f}{N}= \frac{(k_p + \kappa_p) \Delta f}{N} $ and $ \tau_p =\frac{l_{\tau_{p}} T}{M}= \frac{(l_p + \iota_p) T}{M} $, with the corresponding delay and Doppler indices being $ k_{\nu_p} $ and $ l_{\tau_p} $, respectively. Here, $ k_{\nu_p} = k_p + \kappa_p $ and $ l_{\tau_p} = l_p + \iota_p $, where $k_p$ and $ l_p$ denote the Doppler and delay index of the $p$-th path respectively, and the fractional terms  $ \kappa_p, \iota_p \in [-0.5, 0.5] $. The multipath channel is characterized by the channel matrix $ \mathbf{H}[k,l] $, which captures the channel effects at each DDRE. Consequently, the input-output relationship in the DD domain can be expressed as: 

\begin{equation}\label{YDDHDD}
\begin{aligned} 
\mathbf{Y}[k,l] &= \mathbf{H}[k,l] \circledast \mathbf{X}[k,l] + \mathbf{Z}[k,l] \\ 
&= \sum_{p=1}^{P} \alpha_{p} \mathbf{X}[[k - k_{\nu_p}]_N, [l - l_{\tau_p}]_M] + \mathbf{Z}[k, l], 
\end{aligned}
\end{equation}
where $\circledast$ denotes the two-dimensional circular convolution, and $\mathbf{Z} \in \mathbb{C}^{M \times N}$ represents $i.i.d$ complex Gaussian noise. $\mathbb{C}^{M \times N}$ denotes the set of all $M \times N$ matrices with complex entries. $\alpha_{p}$ denotes the channel gain of the $p$-th path, which is related to the reflection and scattering characteristics of the reflectors in the scene. Assume that the transmit pulse and receive pulse satisfy the bi-orthogonal
property, the $\mathbf{H} \in \mathbb{C}^{M \times N}$ can be expressed in \eqref{HDDD}, which is on the top of next page.

\begin{figure*}[t!]
\begin{equation}\label{HDDD}
\begin{aligned}
&\mathbf{H}[k,l]=\sum^{P}_{p=1}\frac{\alpha_{p}e^{j2\pi \frac{k_{\nu_p} l_{\tau_p}}{MN}}}{MN} \underbrace{\left( \sum_{n=0}^{N-1} e^{-j2\pi n \left(\frac{k-k_{\nu_{p}}}{N}\right)} \right)}_{\text{Doppler response}\ \mathbf{h}_{\nu}(k)} \underbrace{\left( \sum_{m=0}^{M-1} e^{j2\pi m \left(\frac{l-l_{\tau_{p}}}{M}\right)} \right)}_{\text{Delay response}\ \mathbf{h}_{\tau}(l)}
\end{aligned}
\end{equation}
\hrule
\end{figure*}

The fractional effects cause signal energy to spread across multiple grid points in the DD domain rather than concentrating at a single grid point, which destroys the sparsity of the channel in the DD domain, leading to a degradation of the channel estimation accuracy. Additionally, the signal energy spreads to adjacent paths, causing IPI, especially when the paths are closely spaced, as illustrated in Fig. \ref{fig_1}\subref{fig_first_case} and Fig. \ref{fig_1}\subref{fig_second_case}.

%

The signal obtained after parallel-to-serial conversion at the receiver is $\mathbf{y}_{\text{vec}} = \text{vec}\{\mathbf{Y}[k,l]\}$. The operator $ \text{vec}\{\cdot\} $performs column-wise vectorization, converting a matrix into a column vector by stacking its columns. Conversely, $ \text{invec}_{M \times N}\{\cdot\} $ reconstructs an $ M \times N $ matrix from a vector by filling its columns sequentially.

\section{Proposed algorithm}
\subsection{Extraction of Path Information}
Consider a single pilot signal transmitted at the $(k=k_{\text{pilot}}, l=l_{\text{pilot}})$-th DDRE. The pilot signal can be expressed as:
\begin{equation}
\begin{aligned}
\mathbf{X}_{\text{pilot}}[k,l]= 
\begin{cases} 
\sqrt{MNE_{p}},&\quad k=k_{\text{pilot}}, l=l_{\text{pilot}} \\ 
0, &\quad otherwise
\end{cases}.
\end{aligned} 
\end{equation}

Upon receiving the signal $\mathbf{y}{\text{vec}}$, it is reshaped into the matrix form $\mathbf{Y}[k,l] = \text{invec}_{M\times N}\{{\mathbf{y}_{\text{vec}}}\}$. The received signal $\mathbf{Y}[k,l]$ contains information from multiple paths in the DDRE. \cite{ref10} proposes a threshold-based estimator for an initial DD domain channel estimate, but it only identifies relatively high-energy paths, potentially omitting low-energy paths and missing potential targets in sensing scenarios.

To address this issue, we improve the method by defining paths in the DDRE as energy peaks in the grid and predefining $P_{\text{max}}$ paths for extraction. The $P_{\text{max}}$ strongest taps in the DD domain are selected, ensuring each tap's energy exceeds its neighbors. Invalid taps are replaced iteratively until $P_{\text{max}}$ valid taps are obtained. Finally, the effective channel $\mathbf{H}$ is recovered using the relationship between $\mathbf{X}_{\text{pilot}}$ and $\mathbf{Y}$ described in \cite{ref10}. The operations of the proposed method are summarized in the Algorithm \ref{al11}.

%
\begin{algorithm}[ht]
\caption{Extraction of Path Information}
\label{alg:path_extraction}
\begin{algorithmic}[1]
\State \textbf{Initialize} $P_{\text{max}}$.
\State Select $P_{\text{max}}$ strongest taps on $\mathbf{Y}$.
\For{each selected tap}
    \If{Energy $\leq$ neighboring cells}
        \State Replace with the next strongest tap.
    \EndIf
\EndFor
\State Recover paths' information using $\mathbf{X}_{\text{pilot}}$ and $\mathbf{Y}$.
\State \Return  Effective channel $\mathbf{H}$ and $P_{\text{max}}$ valid paths.
\end{algorithmic}
\label{al11}
\end{algorithm}

\subsection{Low-complexity Estimation Algorithm}
%

After obtaining the initial estimate of the effective channel, the next goal is to estimate accurately:
\begin{enumerate}
 \item Delay $\tau_{p}={l_{\tau_{p}} T}/{M}$;
 \item Doppler $\nu_p ={k_{\nu_{p}}\Delta f}/{N}$;
 \item Channel gain $\alpha_{p}$.
\end{enumerate}
For the $p$-th path, $p=1,2,\cdots, P_{\text{max}}$, we extract the delay tap $k_{p}$ and Doppler tap $l_{p}$ that correspond to the highest energy in the DDRE. For each path, we then extract the Doppler response $\mathbf{r}_{\nu_{p},l_{p}}(n) = \mathbf{H}[n,l_{p}], n=0,1,\cdots,N-1$ and the delay response $\mathbf{r}_{\tau_{p},k_{p}}(m) = \mathbf{H}[k_{p},m], m=0,1,\cdots,M-1$ from the DDRE. 

This letter leverages the separable relationship between delay and Doppler to directly estimate the delay and Doppler shifts for each path on the effective channel matrix $\mathbf{H}$ in the DD domain. By \eqref{HDDD}, $\mathbf{H}$ can be rewritten as the sum of responses from different paths on the DDRE: 
\begin{equation}\label{hdddd}
\mathbf{H}[k,l] = \sum_{p=1}^{P} \alpha_p \frac{\mathbf{v}_{\tau_p,l}^{T} \mathbf{W}_{\text{\text{in}}} \mathbf{v}_{\nu_p,k}}{MN} \exp\left(-j \frac{2\pi k_{\nu_p} l_{\tau_p}}{MN}\right), 
\end{equation}
where: 
\begin{equation}\label{VV}
\begin{aligned} 
\begin{cases} 
&\mathbf{v}_{\nu_p, k}(n) = \exp\left(-j \frac{2\pi n (k - k_{\nu_p})}{N}\right), \quad n = 0, 1, \dots, N-1, \\ 
&\mathbf{v}_{\tau_p, l}(m) = \exp\left(j \frac{2\pi m (l - l_{\tau_p})}{M}\right), \quad m = 0, 1, \dots, M-1, 
\end{cases} .
\end{aligned} 
\end{equation}
 $\mathbf{v}_{\nu_p, k} \in \mathbb{C}^{N \times 1}$, $\mathbf{v}_{\tau_p, l} \in \mathbb{C}^{M \times 1}$, and $\mathbf{W}_{\text{\text{in}}} \in \mathbb{I}^{M \times N}$, where $\mathbb{I}$ denotes the all-ones matrix.
%

\begin{figure*}[!t]
\centering
\subfloat[]{\includegraphics[width=1.7in]{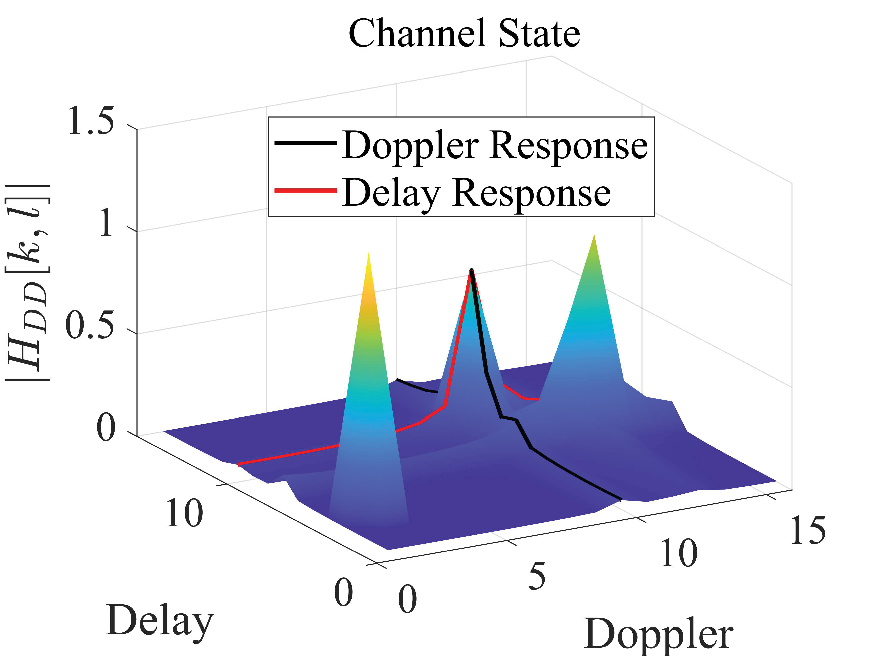}%
\label{fig2a}}
\hfil
\subfloat[]{\includegraphics[width=1.7in]{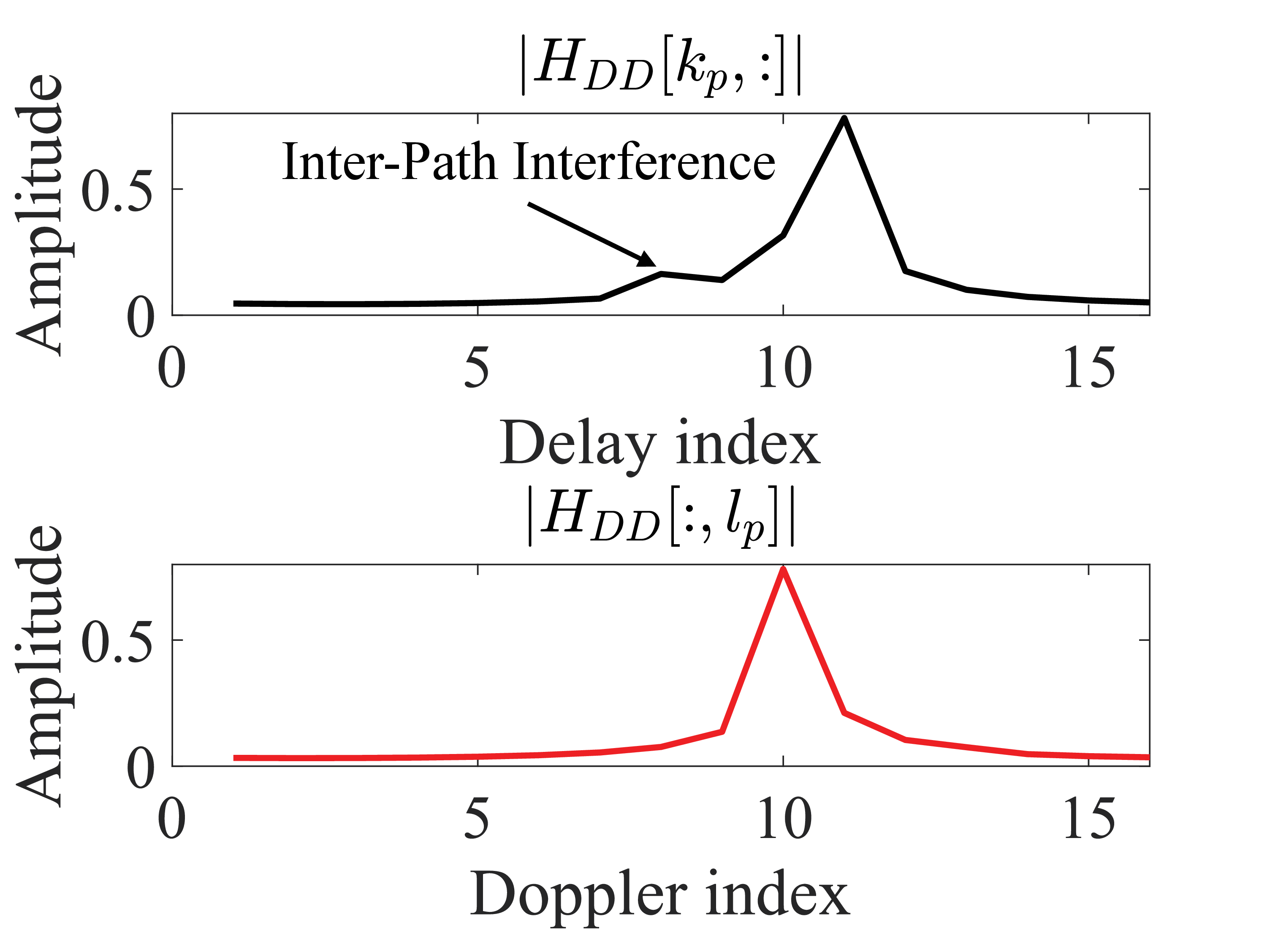}%
\label{fig2b}}
\hfil
\subfloat[]{\includegraphics[width=1.7in]{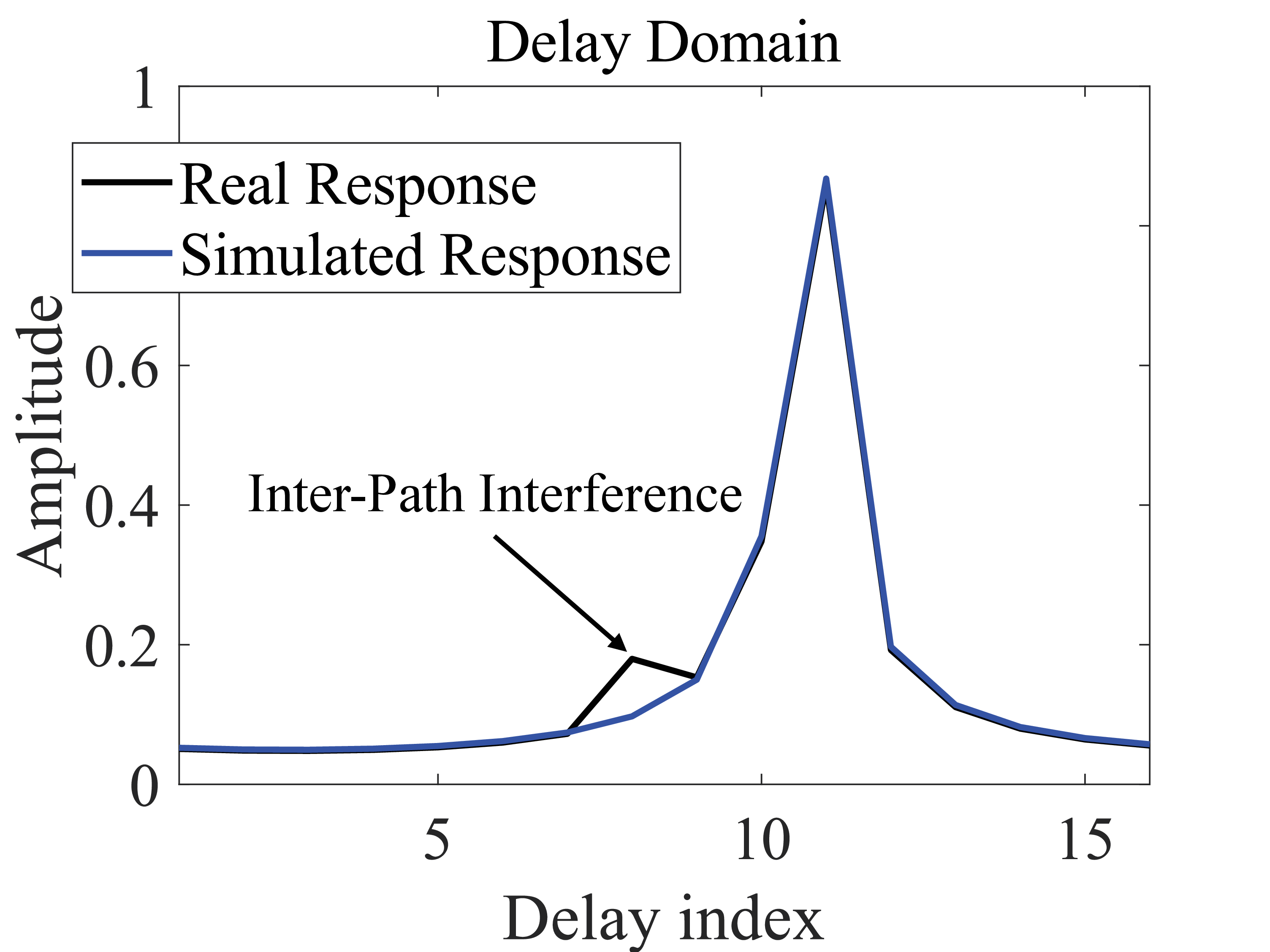}%
\label{fig2c}}
\hfil
\subfloat[]{\includegraphics[width=1.7in]{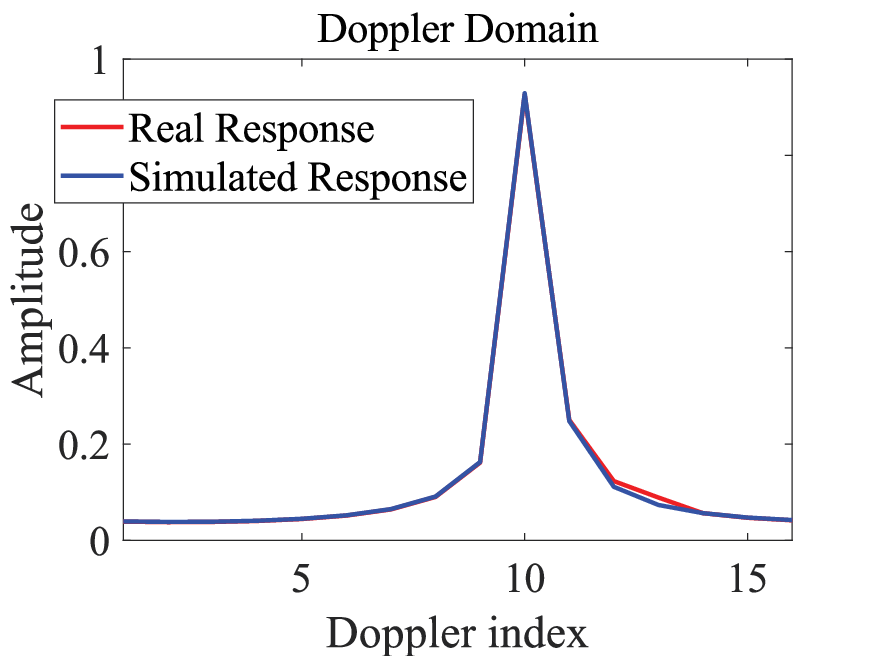}%
\label{fig2d}}
\caption{Schematic representation of the decomposition of the channel state on the DDRE. (a) Channel response on the DDRE. (b) Corresponding Doppler response and delayed response of the path.  (c) The real delay response versus the simulated response calculated by the proposed algorithm.  (d) The real Doppler response versus the simulated response calculated by the proposed algorithm.}
\label{fig2}
\vspace{-10pt} 
\end{figure*}

The channel response of a path can be decomposed into delay and Doppler responses, as shown in Fig. \ref{fig2}\subref{fig2a} and Fig. \ref{fig2}\subref{fig2b}. Therefore, for a given number of DDRE points $N$ and $M$, the simulated delay response $\mathbf{h}_{\tau}(l)\ ( l = 0, 1, \dots, M-1)$ and Doppler response $\mathbf{h}_{\nu}(k)\ (k = 0, 1, \dots, N-1)$ can be calculated, followed by MLE to determine the actual delay and Doppler responses for each path. For the $p$-th path, the search bounds (typically $[k_{p}-1,k_{p}+1]$ and $[l_{p}-1,l_{p}+1]$) are set based on the integer Doppler index $k_{p}$ and delay index $l_{p}$. Considering the Doppler frequency shift estimation range $[\nu_{\text{LB}}, \nu_{\text{UB}}]$ with step size $\nu_{\text{step}}$, the Doppler response is computed for each Doppler frequency value $\nu(i) \in [\nu_{\text{LB}}, \nu_{\text{UB}}], i = 1, 2, \dots, n_{\nu}, n_{\nu}=\frac{\nu_{\text{UB}} - \nu_{\text{LB}}}{\nu_{\text{step}}}$, as follows: 
\begin{equation}\label{dopres}
\mathbf{h}_{\nu}(k) = \sum_{n=0}^{N-1}  \exp\left( j  \frac{2 \pi n(k_{\nu(i)} - k)}{N} \right), \; \forall k = 0, 1, 2, \dots, N-1. 
\end{equation}
Similarly, for the delay estimation range $[\tau_{\text{LB}}, \tau_{\text{UB}}]$ with a step size $\tau_{\text{step}}$, for each delay value $\tau(s) \in [\tau_{\text{LB}}, \tau_{\text{UB}}], \; s = 1, 2, \dots,  m_{\tau}, m_{\tau}=\frac{\tau_{\text{UB}} - \tau_{\text{LB}}}{\tau_{\text{step}}}$, the corresponding delay response is calculated as: 
\begin{equation}\label{delres}
\mathbf{h}_{\tau}(l) = \sum_{m=0}^{M-1} \exp\left( j \frac{2 \pi m (l-l_{\tau(s)})}{M} \right), \; \forall l = 0, 1, 2, \dots, M-1.
\end{equation}
{\eqref{dopres}} and {\eqref{delres}} accurately simulate a path's Doppler or delay domain responses under fixed Doppler or delay conditions. These simulated responses can be multiplied by the channel paths' actual Doppler or delay responses, and their inner product serves as a similarity measure between the simulated and actual responses. Using this similarity, the delay $\tau_p$ and Doppler $\nu_p$ of the $p$-th path can be estimated using the MLE approach, yielding $\hat{\tau}_p$ and $\hat{\nu}_p$ as follows: 

\begin{equation}\label{es1}
\begin{aligned} 
\hat{\nu}_{p} &= \arg \max_{\nu_{p}} \left( |\mathbf{h}_{\nu_p}| \cdot |\mathbf{r}_{\nu_p, l_p}| \right), \\ 
\hat{\tau}_{p} &= \arg \max_{\tau_{p}} \left( |\mathbf{h}_{\tau_p}| \cdot |\mathbf{r}_{\tau_p, k_p}| \right). 
\end{aligned}
\end{equation}

After estimating the Doppler and delay values of each path, the channel coefficient for the $i$-th path, $\hat{\alpha}_i$, can be estimated using \eqref{hdddd} as: 
\begin{equation}\label{es2}
\hat{\alpha}_{p} = \frac{MN \mathbf{H}[k_{p}, l_{p}]}{\mathbf{v}_{\hat{\nu}_p, l}^T \mathbf{W}_{\text{\text{in}}} \mathbf{v}_{\hat{\tau}_p, k} \exp\left( -j \frac{2\pi \hat{k}_{\nu_p} \hat{l}_{\tau_p}}{MN} \right)}.
\end{equation}

\subsection{Sequential Estimation and IPI Elimination}

Energy leakage caused by fractional effects introduces interference among paths, making it essential to eliminate the interference progressively for more accurate path estimation. Paths with severe fractional effects exhibit more pronounced energy leakage in the DDRE, causing more significant interference with neighboring paths. Thus, it is necessary to quantify the severity of fractional effects for each path. The energy leakage level of a path is defined as the reciprocal of the ratio of its maximum energy at the delay and Doppler position $|\mathbf{H}[k_p, l_p]|$ to the total energy of its neighboring elements. Specifically, the energy leakage level of the $p$-th path is defined as follows:
\begin{equation}
\mathscr{L}(p) = \frac{\sum_{(k, l) \in \text{Neighbors}(k_p, l_p)} |\mathbf{H}[k, l]|}{|\mathbf{H}[k_p, l_p]|},
\end{equation}
where $\text{Neighbors}(k_p, l_p)$ represents the set of neighboring elements of the path position $(k_p, l_p)$, defined as: 
\begin{equation}
\begin{aligned}
\text{Neighbors}(k_p, l_p) = \{\mathbf{H}[[k_p - 1]_{N} , l_p], \, \mathbf{H}[[k_p + 1]_{N} , l_p],\\
 \, \mathbf{H}[k_p, [l_p - 1]_{M}], \, \mathbf{H}[k_p, [l_p + 1]_{M}\}, 
\end{aligned}
\end{equation}
where $(k_p, l_p)$ corresponds to the maximum integer Doppler/delay position of the $p$-th path in the DDRE, the operations $[\cdot]_M$ and $[\cdot]_N$ denote modular arithmetic in $M$ and $N$, respectively, satisfying the boundary conditions for cyclic structures. 

The path with the most severe fractional effects is estimated firstly, yielding its delay, Doppler information $\hat{\tau}_{p}$, $\hat{\nu}_{p}$, and channel coefficient $\hat{\alpha}_{p}$. Using this information, the channel containing only the $p$-th path $\mathbf{H}_{p}[k, l]$ can be reconstructed by \eqref{hdddd}. The reconstructed channel is then subtracted from the known channel and continues to estimate the $(p+1)$-th path until the $P_{\max}$-th path has been estimated. The operations of the proposed method are summarized in the Algorithm \ref{al1}.

\subsection{The Complexity Analysis of Proposed Algorithm}
The operations of the proposed algorithm are shown in the Table \ref{complexity}. The asymptotic complexity of the proposed algorithm after omitting the constant factor and lower order terms is $O(MN\log(MN) + m_{\tau}M^{2} + n_{\nu}N^{2} + P_{\max}(m_{\tau}M + n_{\nu}N + MN))$. Since the proposed algorithm does not require iterations, the complexity is significantly lower than the low-complexity algorithm proposed in \cite{ref8}.

%
%

\begin{algorithm}[ht]
\caption{Sequential Path Estimation and IPI Elimination}
\label{alg:seq_path_estimation}
\begin{algorithmic}[1]
\State \textbf{Initialize}: Set the maximum number of paths $P_{\text{max}}$.
\State \textbf{Evaluate and Rank Leakage Levels}: Compute the leakage level $\mathscr{L}_p$ for all paths and sort them in descending order to form the leakage level list $\{\mathscr{L}_1, \mathscr{L}_2, \mathscr{L}_3, \dots, \mathscr{L}_{P_{\text{max}}}\}$. Set $p = 1$.

\While{$p \leq P_{\text{max}}$}
    \State \textbf{Path Selection}: Select the path with the highest leakage level from the list. For the $p$-th path, extract $\mathbf{r}_{\nu_p, l_p}$, $\mathbf{r}_{\tau_p, k_p}$, and $\mathbf{H}(k_p, l_p)$. Estimate $\hat{\tau}_p$, $\hat{\nu}_p$, and $\hat{\alpha}_p$ by \eqref{es1} and \eqref{es2}.
    
    \State \textbf{Channel Reconstruction}: Using $\hat{\tau}_p$, $\hat{\nu}_p$, and $\hat{\alpha}_p$, reconstruct the channel containing only the $p$-th path, $\mathbf{H}_{p}$.
    
    \State \textbf{Interference Elimination}: Subtract the reconstructed channel from the known channel to remove the influence of the $p$-th path:
    \[
    \mathbf{H}[k, l] = \mathbf{H}[k, l] - \mathbf{H}_{p}[k, l].
    \]
    
    \State \textbf{Update}: $p = p + 1$.
\EndWhile 

\State \textbf{Output}: Output the estimated parameters of all paths as:
\[
\{[\hat{\tau}_1, \hat{\nu}_1, \hat{\alpha}_1], [\hat{\tau}_2, \hat{\nu}_2, \hat{\alpha}_2], \dots, [\hat{\tau}_{P_{\text{max}}}, \hat{\nu}_{P_{\text{max}}}, \hat{\alpha}_{P_{\text{max}}}]\}.
\]

\end{algorithmic}
\label{al1}
\end{algorithm}

\begin{table*}[]
\centering
\caption{Approximate Complexity Analysis of Operations.}
\begin{tabular}{cccc}
\hline
Operations                       & Complex Multiplications          & Addition Complex                         & \textbf{Total Complex}                    \\ \hline
Finding $P_{\max}$ paths.         & \multicolumn{2}{c}{\bf{N/A}(Comparisons Only)}                                   & $MN\log(MN)$                              \\
Leakage Calculation and Sorting. & $10P_{\max}$                      & $8P_{\max}$                               & $18P_{\max}$                               \\
Generation of Delay response.    & $m_{\tau} M^{2}$                 & $m_{\tau} M^{2}$                         & $2m_{\tau} \cdot M^{2}$                   \\
Generation of Doppler response.  & $n_{\nu} N^{2}$                  & $n_{\nu}N^{2}$                           & $2n_{\nu}N^{2}$                           \\
Estimation of $\tau$ and $\nu$.  & $P_{\max}(m_{\tau}M + n_{\nu}N)$ & $P_{\max}[m_{\tau}(M-1) + n_{\nu}(N-1)]$ & $\approx 2P_{\max}[m_{\tau}M + n_{\nu}N]$ \\
Estimation of $\alpha$.          & $P_{\max} MN$                    & $P_{\max} MN$                            & $2 P_{\max} MN$                           \\ \hline
\end{tabular}
\label{complexity}
\vspace{-10pt} 
\end{table*}

\section{Numerial Results}
\subsubsection{Simulation Setup}

The simulation uses DDRE parameters of $ M = 64 $, $ N = 32 $, $ \Delta f = 30 $ kHz, and $ f_c = 5.1 $ GHz, with five paths, including one line-of-sight path and a Rice factor of 15 dB\cite{ref8}. The delay between the first and second paths is set to $ 0.2 \, \mu s $, and the delays of the other paths are uniformly distributed between $ 0.867 \, \mu s $ and $ 7 \, \mu s $. Doppler frequencies for all paths are generated based on Jake's Doppler spectrum, with $ \nu_p = \nu_{\text{max}} \cos(\theta_p) $, where $ \theta_p $ is uniformly distributed in $ (0, 2\pi] $. The maximum Doppler shift is set to 1700 Hz to emulate dynamic effects.

Increasing $ P_{\text{max}} $ in channel estimation improves the extraction of channel details and estimation performance. A higher $ P_{\text{max}} $ extracts more path information, enhancing accuracy. In \cite{ref9}, $ P_{\text{max}} $ was set larger than the actual number of paths to ensure accuracy in complex channels, which increased computational complexity. To reduce the computational load and improve efficiency, the letter sets $ P_{\text{max}}= 5 $. $\tau_{\text{step}}, \nu_{\text{step}}=0.01$. The algorithm's evaluation metrics are as follows: for channel estimation, the Normalized Mean Square Error ($\text{NMSE}=10\log_{10}[\frac{||\hat{\mathbf{H}}-\mathbf{H}||^{2}_{2}}{||\mathbf{H}||^{2}_{2}}]$) measures channel estimation performance. Pilot signal-to-noise ratio (PSNR) is defined as the ratio of the average power of the transmitted time-domain pilot signal to the AWGN power at the receiver \cite{ref8}. The letter uses a blank pilot block at the transmitter and receiver as \cite{ref9}, so the $\text{PSNR}=\frac{E_{p}}{MN}$. For radar parameter sensing, the Mean Square Error (MSE) of range and velocity measurements is used to assess performance as $\text{PSNR}$ varies. The M-MLE algorithm in \cite{ref8} outperforms other algorithms, while the DDIPIC algorithm in \cite{ref9} enhances M-MLE for even better performance. Thus, this letter compares the method in \cite{ref8} and \cite{ref9}.

\subsubsection{Results discussion}
\begin{figure}[!t]
\centering

\subfloat[\fontsize{1}{1}\selectfont]{\includegraphics[width=1.72in]{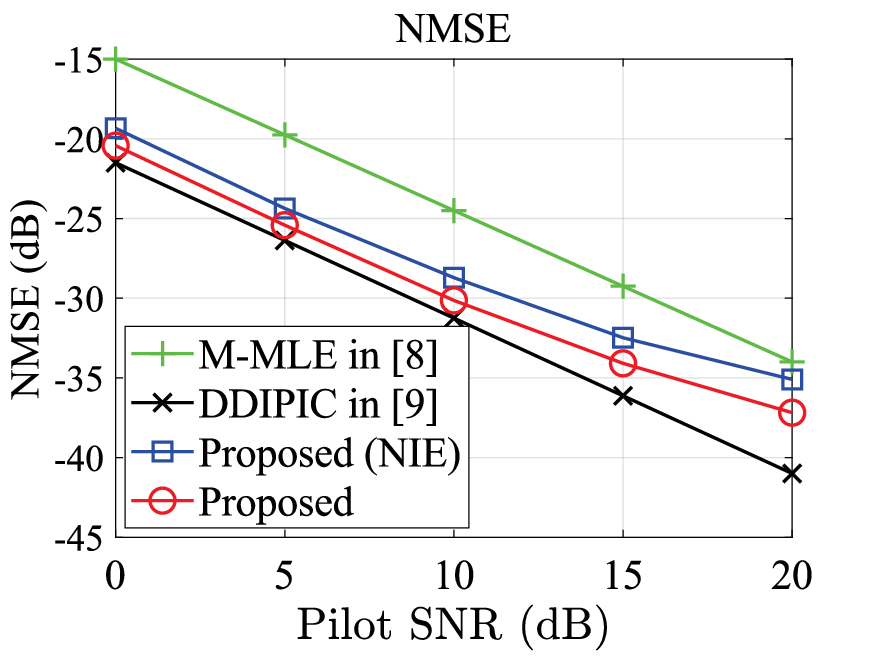}\label{fig5a}}
\hfil
\subfloat[\fontsize{1}{1}\selectfont]{\includegraphics[width=1.72in]{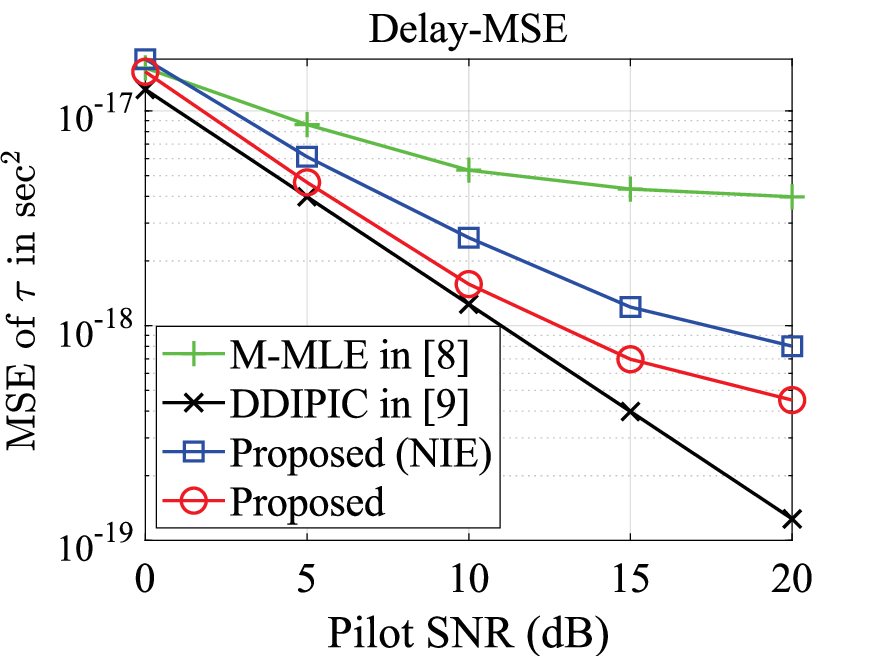}\label{fig5b}}
\hfil
\subfloat[\fontsize{1}{1}\selectfont]{\includegraphics[width=1.72in]{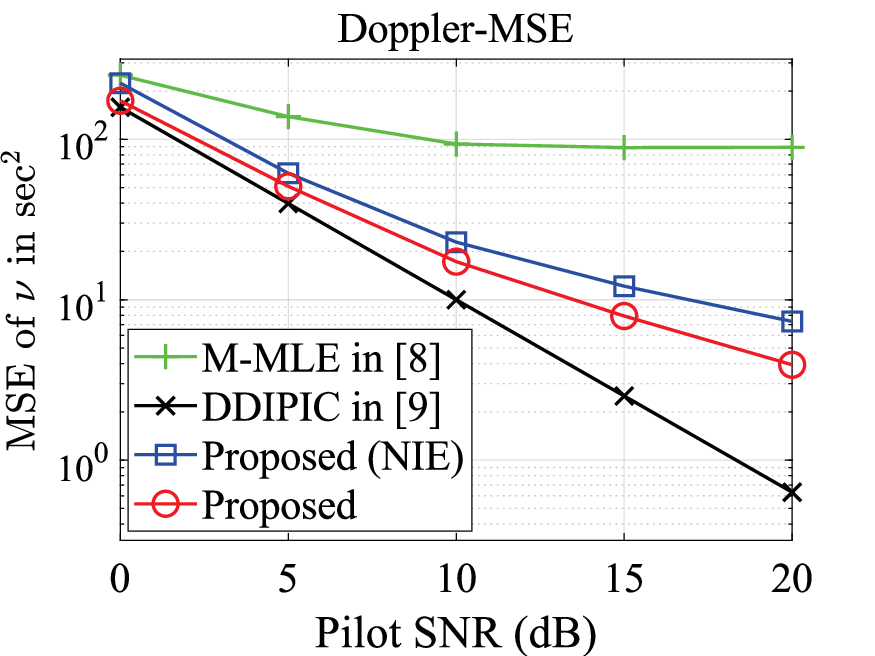}\label{fig5c}}
\hfil
\subfloat[\fontsize{1}{1}\selectfont]{\includegraphics[width=1.72in]{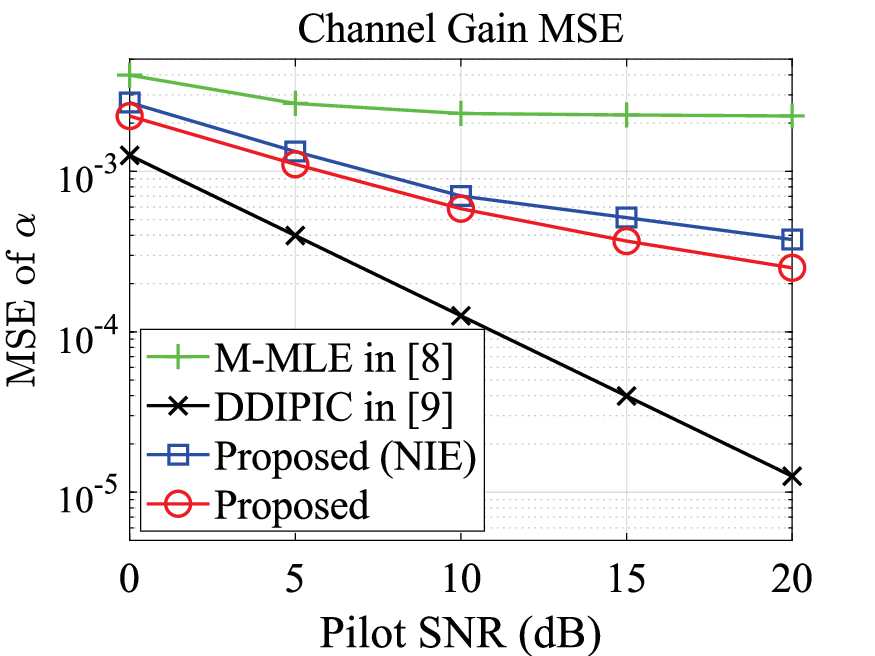}\label{fig5d}}
\hfil
\subfloat[\fontsize{1}{1}\selectfont]{\includegraphics[width=2.12in]{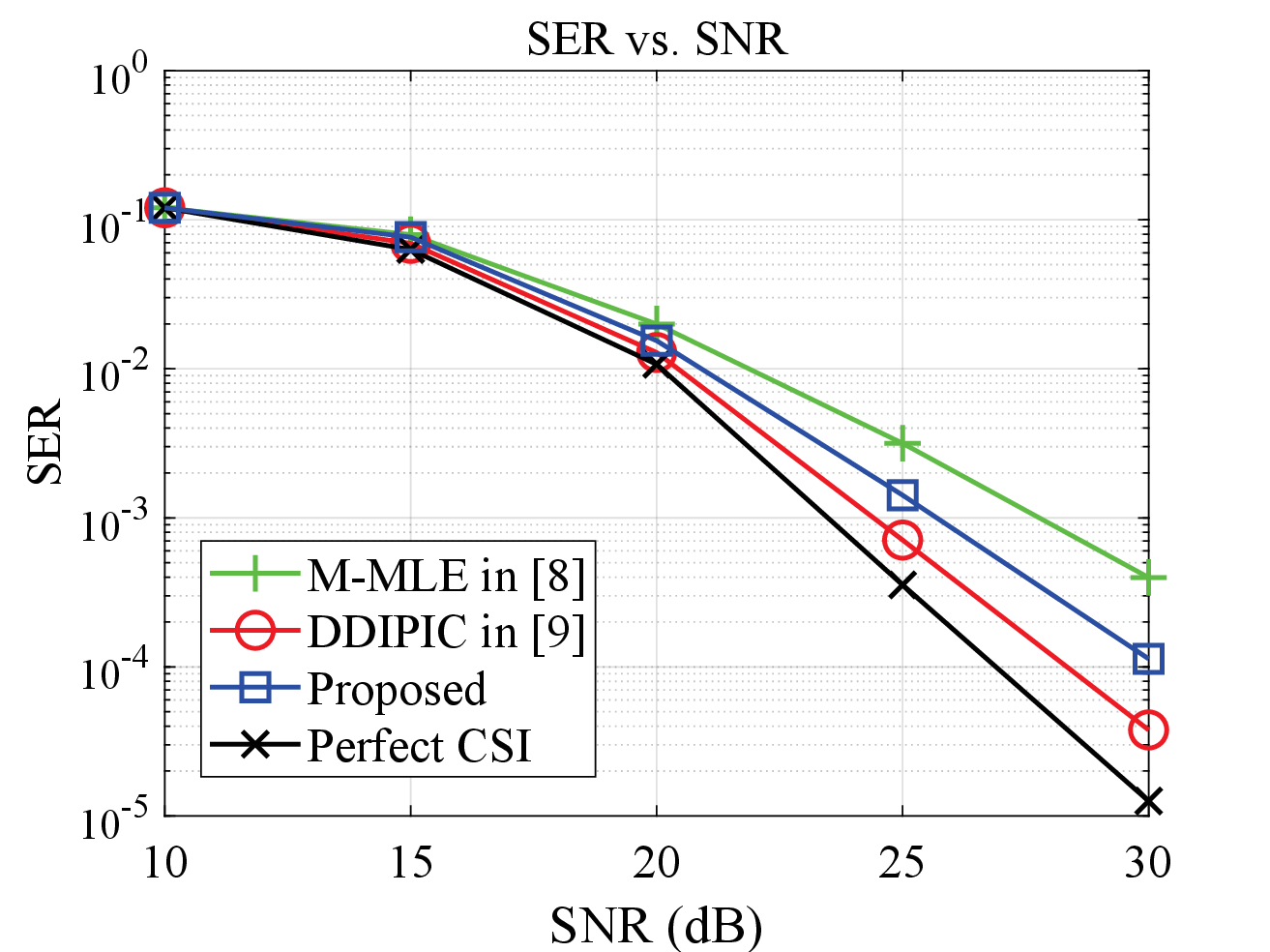}\label{fig6}}
\caption{(a) NMSE performance of the proposed algorithm as a function of pilot SNR. (b) MSE of estimated Doppler as a function of pilot SNR.  (c) MSE of estimated delay as a function of pilot SNR. (d) MSE of estimated Channel gain as a function of pilot SNR. (e) SER performance of the proposed method in comparison with other methods.}
\label{fig_3}
\vspace{-15pt} 
\end{figure}

Fig. \ref{fig_3} illustrates the performance of the M-MLE, the DDIPIC algorithm, the proposed algorithm with No IPI elimination (NIE), and the proposed algorithm with IPI elimination in terms of channel estimation and assisting sensing, which provides the NMSE of channel estimation and MSE for estimating delay, Doppler and channel gain.
 
The results demonstrate that the NMSE of the proposed algorithm decreases approximately linearly with increasing PSNR. By incorporating an IPI elimination mechanism, the proposed algorithm significantly enhances the robustness and accuracy of channel estimation in scenarios with severe IPI, as validated through comparisons with the NIE curve. As shown in Fig. \ref{fig_3}\subref{fig5a}, the proposed algorithm outperforms the M-MLE algorithm in channel estimation performance and achieves nearly comparable performance to the DDIPIC algorithm at low PSNR levels. Moreover, the proposed algorithm requires no iterative processes, resulting in significantly reduced complexity compared to the M-MLE algorithm and far lower computational complexity than the DDIPIC algorithm, highlighting its efficiency. 

Fig. \ref{fig_3}\subref{fig5b} and Fig. \ref{fig_3}\subref{fig5c} compare the MSE curves for delay and Doppler estimation. The results indicate that the proposed algorithm achieves high accuracy in estimating delay and Doppler parameters, enabling effective perception of dynamic parameters such as distance and velocity in the environment. 
However, Fig. \ref{fig_3}\subref{fig5d} reveals that the proposed algorithm's MSE of channel gain estimation is less favorable. This limitation is attributed to the assumption in \eqref{es2} that the value of $\mathbf{H}[k_{p},l_{p}]$ is solely determined by the parameters of the $p$-th path, neglecting the potential impact of other paths. To enhance the accuracy of channel gain estimation, future work can focus on introducing simple iterative schemes for more accurate channel gain estimation.

Fig. \ref{fig_3}\subref{fig6} shows the Symbol Error Rate (SER) versus Signal-to-Noise Ratio (SNR). The proposed algorithm reduces SER with increasing SNR, closely following the perfect Channel State Information (CSI) trend, indicating effective channel reconstruction through accurate delay, Doppler shift, and channel gain estimation. Although slightly less effective than the DDIPIC algorithm, removing the iterative process significantly reduces computational complexity. Compared to the methods in \cite{ref8} and \cite{ref9}, it offers efficient and accurate channel estimation with much lower complexity.

%

\section{Conclusion}
%

The letter proposes a low-complexity OTFS channel estimation method under the fractional effect. A novel path identification method is introduced by leveraging the energy characteristics of paths in the DD domain. By modeling delay and Doppler responses, accurate estimation of path's delay and Doppler shift is achieved. Iterative subtraction of the reconstructed channel effectively eliminates IPI. Experimental results validate the effectiveness of the proposed algorithm.
%
{

}


\bibliographystyle{IEEEtran}

\bibliography{IEEEfull}
\

\end{document}